\documentclass{aa}

\usepackage{graphicx,natbib}
\usepackage[varg]{txfonts}

\bibpunct{(}{)}{;}{a}{}{,} 

\begin{document}

\title{Hydrodynamical simulations for the common-envelope wind model for Type Ia supernovae}

\author{Yingzhen Cui\inst{1,2}\and Xiangcun Meng\inst{1,3,4}\and Philipp Podsiadlowski\inst{5}\and Ren Song\inst{1,2}}

\institute{Yunnan Observatories, Chinese Academy of Sciences, Kunming 650216, P.R. China
\and University of Chinese Academy of Sciences, Beijing 100049, P.R. China\\e-mail: cuiyingzhen@ynao.ac.cn
\and Key Laboratory for the Structure and Evolution of Celestial Objects, Chinese Academy of Sciences, Kunming 650216, P.R. China
\and Center for Astronomical Mega-Science, Chinese Academy of Sciences, 20A Datun Road, Chaoyang District, Beijing 100012, P.R. China\\e-mail: xiangcunmeng@ynao.ac.cn
\and University of Oxford, St Edmund Hall, Oxford, OX1 4AR, UK\\e-mail: podsi@hotmail.com}

\abstract{The single-degenerate (SD) model is one of the leading models for the progenitors of Type Ia supernovae (SNe Ia). Recently, a new version of the SD model, the common-envelope wind (CEW) model, has been proposed, which, in principle, has the potential to resolve most of the difficulties encountered by previous SD models. This model is still being developed and a number of open issues remain, such as the details of the mass-loss mechanism from the surface of the common envelope (CE), the main observational properties, and the spiral-in timescale of the binary inside the envelope. }
{In this article, we aim to address these issues by considering hydrodynamical effects on the CE.}
{Using the stellar evolution code MESA, we carried out a series of 1D hydrodynamical simulations of an asymptotic giant branch (AGB) star undergoing a common-envelope phase with different envelope masses (0.0007$\,\mathrm{M}_\sun$-0.06$\,\mathrm{M}_\sun$). The effect of the immersed binary was mimicked by changing the gravitational constant throughout the envelope and injecting an extra heating source at the location of the binary orbit. }
{We found that the envelopes are always dynamically unstable, leading to regular mass ejection events if the envelope is more massive than the critical value of $\sim 0.003\,\mathrm{M}_\sun$. The $\kappa$ mechanism can naturally explain this phenomenon. We also found that, due to the low mass of the CE, the estimated frictional luminosity caused by the spiral-in of the immersed binary is much less than the nuclear luminosity, and therefore will not affect the structure of the CE significantly.}
{Our results imply that the CE in the CEW model cannot be very massive. We also present a rough estimate for the spiral-in timescale based on a simplified model. We found that, for reasonable assumptions, the timescale may be longer than a few $10^5$ yr; therefore, the white dwarf (WD) may have enough time to increase its mass toward the Chandrasekhar mass, avoiding a merger with the companion.}


\titlerunning{Hydrodynamic simulation for CEW model of SNe Ia}
\authorrunning{Cui et al.}

\maketitle

\section{Introduction}
Type Ia supernovae (SNe Ia) are some of the most energetic events in the Universe and play an important role in many astrophysical fields. They are used as cosmological distance indicators, which has led to the discovery of the accelerating expansion of the Universe \citep{Riess1998AJ,Perlmjtter1999ApJ}. It is widely accepted that a SN Ia originates from a thermonuclear runaway of a carbon-oxygen white dwarf (CO WD) in a binary system \citep{Hillebrandt2013FrPhy}, where carbon is ignited
under degenerate conditions when the WD reaches a critical mass.
The main problem for understanding the progenitors of SNe Ia is how the WD grows in mass to reach the critical mass for explosion, that is,\ the Chandrasekhar mass ($M_{\rm ch}$).

At present, two main scenarios are being debated for the progenitor systems of SNe Ia. One is the single-degenerate (SD) model in which the WD accretes material from a nondegenerate companion \citep{Whelan1973ApJ,Nomoto1984ApJ}. The other is the double-degenerate (DD) model, which most commonly involves the merger of two CO WDs \citep{Iben1984ApJS,Webbink1984ApJ}. In this paper, we focus on the SD model.

In the canonical SD model, a CO WD accretes H-rich or He-rich materials from its companion star and explodes as a SN Ia once its mass finally approaches $M_{\rm ch}$. Mass transfer between the CO WD and its companion may be dynamically stable or unstable; this depends on the mass ratio of the donor to the accreting WD and the evolutionary state of the companion when it fills its Roche lobe \citep{Hjellming1987ApJ,Ge2015ApJ...812...40G}. For example, if the companion fills its Roche lobe on the main sequence, unstable mass transfer will occur for a mass ratio larger than approximately four \citep{Han2000MNRAS,Chen2002MNRAS,Chen2003MNRAS}. Unstable mass transfer will lead to the formation of a common envelope (CE), and the system may then quickly merge instead of producing a SN Ia. For dynamically stable mass transfer, there is a critical accretion rate for the accreting WD. If the mass-transfer rate exceeds the critical accretion rate, the WD will expand and become a giant-like object and finally engulf the companion; in other words, a CE can form again \citep{2007ApJ...663.1269N}. At present, the subsequent evolution of the system with this type of envelope is unclear. It is often assumed that the binary will lose its orbital angular momentum and spiral in quickly because of the frictional drag caused by the CE. Eventually, such a binary will merge, again not producing a SN Ia \citep{Nomoto1979PASJ}. If the mass transfer onto the WD is too low, nova explosions and helium flashes may eject most of the accreted matter, making it difficult for the WD to grow in mass. Because of this fine-tuning required for the mass-transfer rate, the birth rate of SNe Ia from the SD model appears to be too low to be compatible with observations. 

In order to overcome this problem, \citet{Hachisu1996ApJ} proposed an optically thick wind (OTW) model, in which the OTW regulates the mass transfer to prevent the formation of a CE. As a result, the birth rate of SNe Ia from the OTW model may be comparable with the observationally inferred rate \citep{Han2004MNRAS,Meng2009MNRAS.395.2103M}. However, some predictions from the OTW model are in conflict with observations. For example, the wind velocity of the OTW normally exceeds 1000\,km/s, which is much higher than that deduced observationally \citep{Badenes2007ApJ}. Meanwhile, \citet{Kobayashi1998ApJ} pointed out that the OTW cannot occur if $ Z\leq0.002 $, which means that the OTW model cannot explain SNe Ia in low-metallicity and/or high-redshift environments \citep{Frederiksen2012ApJ,Rodney2012ApJ...746....5R,Jones2013ApJ...768..166J,Rodney2015AJ....150..156R}.

The above contradictions imply that there should not be an OTW phase during the pre-explosion evolution of a typical SN Ia. To overcome some of these problems, \citet{Meng2017MNRAS} proposed a new version of the SD model, which they referred to as the ``common-envelope wind (CEW) model''. The model postulates that if the mass-transfer rate between a CO WD and its main-sequence (MS) companion exceeds the critical accretion rate, the WD will expand to giant dimensions, leading to the formation of a CE around the binary system instead of an OTW. The WD will then gradually increase its mass at the base of the CE, a situation that is similar to what is encountered in thermally pulsing asymptotic branch (TPAGB) stars. Because of the low density of the CE, the binary may avoid a fast spiral-in phase and finally survive from the CE phase. \citet{Meng2017MNRAS} found that the birth rate of SNe Ia from the CEW model is about 30\% higher than that from the OTW model. Based on the CEW model, \citet{Meng2018ApJ} found that SN 2002cx-like and SN Ia-CSM\footnote{SNe Ia showing a strong circumstellar medium (CSM) signal are classified as SNe Ia-CSM \citep{Silverman2013ApJS..207....3S}.} could share the same origin. \citet{Meng2019MNRAS} studied the properties of the companions at the moment of the supernova explosion based on the CEW model and found that the surviving companions of some SN Ia could be subdwarf B stars. In particular, \cite{Meng2020ApJ...903..100M} suggested that the mysterious variables, the so-called blue large-amplitude pulsators (BLAPs), could be the long-sought surviving companions of SNe Ia.

However, at present the CEW model is still under development, and there are quite a few uncertainties remaining. First, for simplicity, \citet{Meng2017MNRAS} applied a modified Reimer's wind to calculate the mass-loss rate, but the detailed mechanism for the mass loss is actually quite unclear. Second, the spiral-in timescale of the binary system in the CE is also uncertain. \citet{Meng2017MNRAS} simply used an average density of the CE to estimate the spiral-in timescale and found that it can be longer than $ 10^6 $ yr for most cases, which is long enough to permit the WD to increase its mass to $ M_\mathrm {ch} $. However, \citet{song2020AA} recently re-estimated the spiral-in timescale by using a density distribution obtained from 1D hydrostatic simulations, and found that the binary will merge within several hundred years if the envelope is as massive as 0.6$ \,\mathrm{M}_\sun $. This contradiction is due to the fact that the envelopes in the model of \citet{song2020AA} were too massive and that hydrodynamical effects were neglected. In particular, the cases with massive CEs in \citet{Meng2017MNRAS} were not commonly found, with typical CE masses in the range of $10^{-4}-10^{-3}\,\mathrm{M}_\sun$, assuming a modified Reimer's wind. Finally, according to the 1D hydrostatic simulations in \citet{song2020AA}, the systems would still look similar to canonical TPAGB stars if the CEs were very massive. However, it is not clear what such systems would look like during the CE phase if the CE mass was only in the range of $10^{-4}-10^{-3}\,\mathrm{M}_\sun$ and if hydrodynamical effects were being considered.

In this paper, we present hydrodynamical simulations of the properties of systems with a low-mass CE, using the model of  \citet{song2020AA}, to address the abovementioned problems of the CEW model. In Sect.~2, we describe the numerical methods for the simulations, and we present the results in Sect.~3. We discuss our results in Sect.~4, and summarize them in Sect.~5. 

\section{Method}
In this study, we used the Modules for Experiments in Stellar Astrophysics (MESA) code \citep[version 10398;][]  {Paxton2011ApJS,Paxton2013ApJS,Paxton2015ApJS,Paxton2018ApJS,Paxton2019ApJS} to carry out 1D hydrodynamical simulations for systems with a low-mass CE in the CEW model. Similar to \citet{song2020AA}, we used a modified TPAGB star to represent the binary system in the CE. The effects of the companion's gravity and the rotation of the CE were simulated by modifying the gravitational constant, and the energy from the friction between the binary and the CE was treated as an extra heating source \citep[for details see][]{song2020AA}.

We constructed the initial models according to the following steps. Since the structure of the system during the CE phase should be similar to a TPAGB star, we used a TPAGB star with an initial mass of 7.2$\,\mathrm{M}_\sun$ to construct our initial models. When the CO core of the TPAGB reached a mass of 1$\,\mathrm{M}_\sun$, we added the modifications described in \citet{song2020AA} to alter the structure of the star, to simulate the effects of the immersed binary. We then reduced the hydrogen-rich envelope mass artificially,
at a rate of $10^{-2}\,\mathrm{M}_\sun\,\mbox{yr}^{-1}$ until the envelope mass was between $0.0007\,\mathrm{M}_\sun$ and $0.06\,\mathrm{M}_\sun$. Because of the rapid mass loss, the star would be out of thermal equilibrium. Then, after the mass-loss routine was turned off, we allowed the star to relax for 10 yr, which was at least one order of magnitude longer than the thermal timescale of the envelope. Using this process, we obtained a series of initial models with different envelope masses. At this point, we activated the hydrodynamics option in MESA to study the dynamical effects. 

In all of our simulations, the metallicity was 0.02 and the ratio of mixing length to local pressure scale height, $ \alpha = l/H_\mathrm{p} $, was set to two \citep{Pols1997MNRAS.289..869P}. We adopted the Eddington gray atmosphere model. The mass and the luminosity of the companion star was set to $1\,\mathrm{M}_\sun $  and $ 1\,\mathrm{L}_\sun  $ for simplicity. The initial separation $ a $ was $40\,\mathrm{R}_\sun$. The parameters were specifically chosen to match those in \cite{song2020AA}, but we found in test calculcations that the main results were not very sensitive to the exact parameters. The only adjustable parameter was the envelope mass in our simulations. The main purpose of this study was to determine how the envelope mass affects the mass loss and estimate the mass-loss rate.  

As we show in the next section, based on the hydrodynamical simulations, the velocity of the surface material for some systems will exceed the surface escape velocity and this unbound material will be ejected from the systems. Here, we used the method proposed by \citet{Clayton2017MN} to remove the unbound material by $ other\_wind $ routine in MESA. In order to minimize the influence of the mass removal process on the simulations, we set a mass-loss rate of 1000$\,m\,\mathrm{yr}^{-1}$, where $ m $ is the mass of the layer that becomes unbound in each time step. This mass-loss rate implies that the unbound material is ejected within $ 10^{-3} $ yr, which is at least one order of magnitude shorter than the dynamical and thermal timescales \citep[see detail in][]{Clayton2017MN}.

\begin{figure*}
        \centering
        \includegraphics[width=17cm]{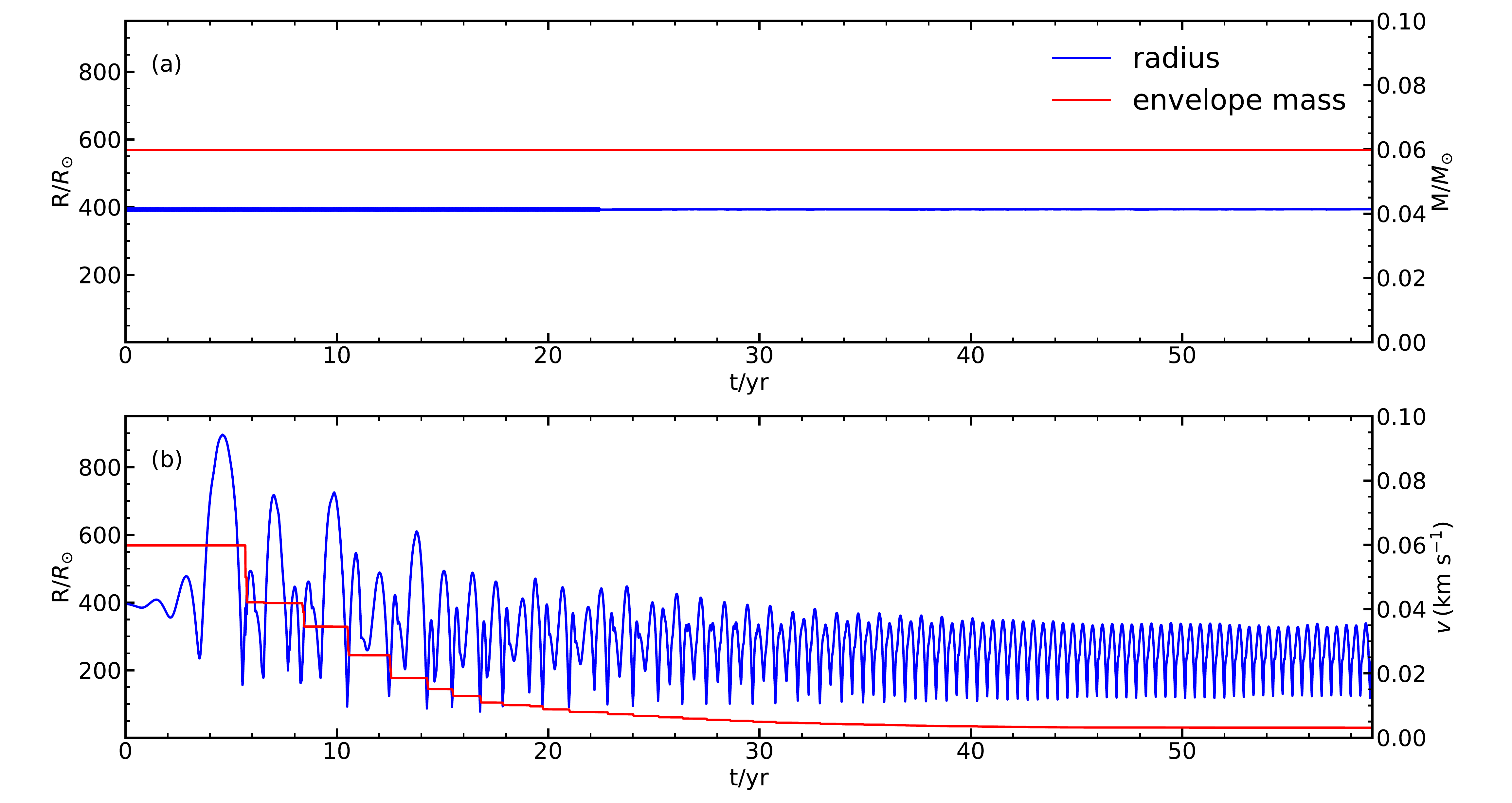}
        \caption{Evolution of the surface radius (blue) and envelope mass (red) for the model with an envelope of $ 0.06\,\mathrm{M}_\sun $. Panel (a) shows the result of the hydrostatic simulation and panel (b) the result of the hydrodynamical simulation.}
        \label{different methods}
\end{figure*} 

\section{Results}
\subsection{Why a hydrodynamical treatment is required}
Our simulations show that the systems considered here are always dynamically unstable and that hydrostatic simulations are not adequate. As an example, Fig.~\ref{different methods} shows the evolution of the radius and the envelope mass of the hydrostatic and hydrodynamical simulations of a model with an initial envelope mass of $0.06\,\mathrm{M}_\sun$. In the hydrostatic simulation, as shown in Fig.~\ref{different methods}(a), the system is in hydrostatic equilibrium; in other words, the radius of the system does not change during the evolution. However, an important question is whether the equilibrium is dynamically stable, which can be examined by testing the response of a system to a dynamical perturbation by applying a hydrodynamical treatment; in other words, a small perturbation will lead to a pulsation with a large amplitude if the system is dynamically unstable, or otherwise the perturbation will be quickly dissipated. As shown in Fig.~\ref{different methods}(b), when we activate the hydrodynamical option in the simulation, the system becomes unstable and begins to pulsate with a rapidly increasing amplitude. This shows that a hydrodynamical treatment is necessary to understand the evolution of the systems during the CE phase in the CEW model. 

\begin{figure*}
        \centering
        \includegraphics[width=17cm]{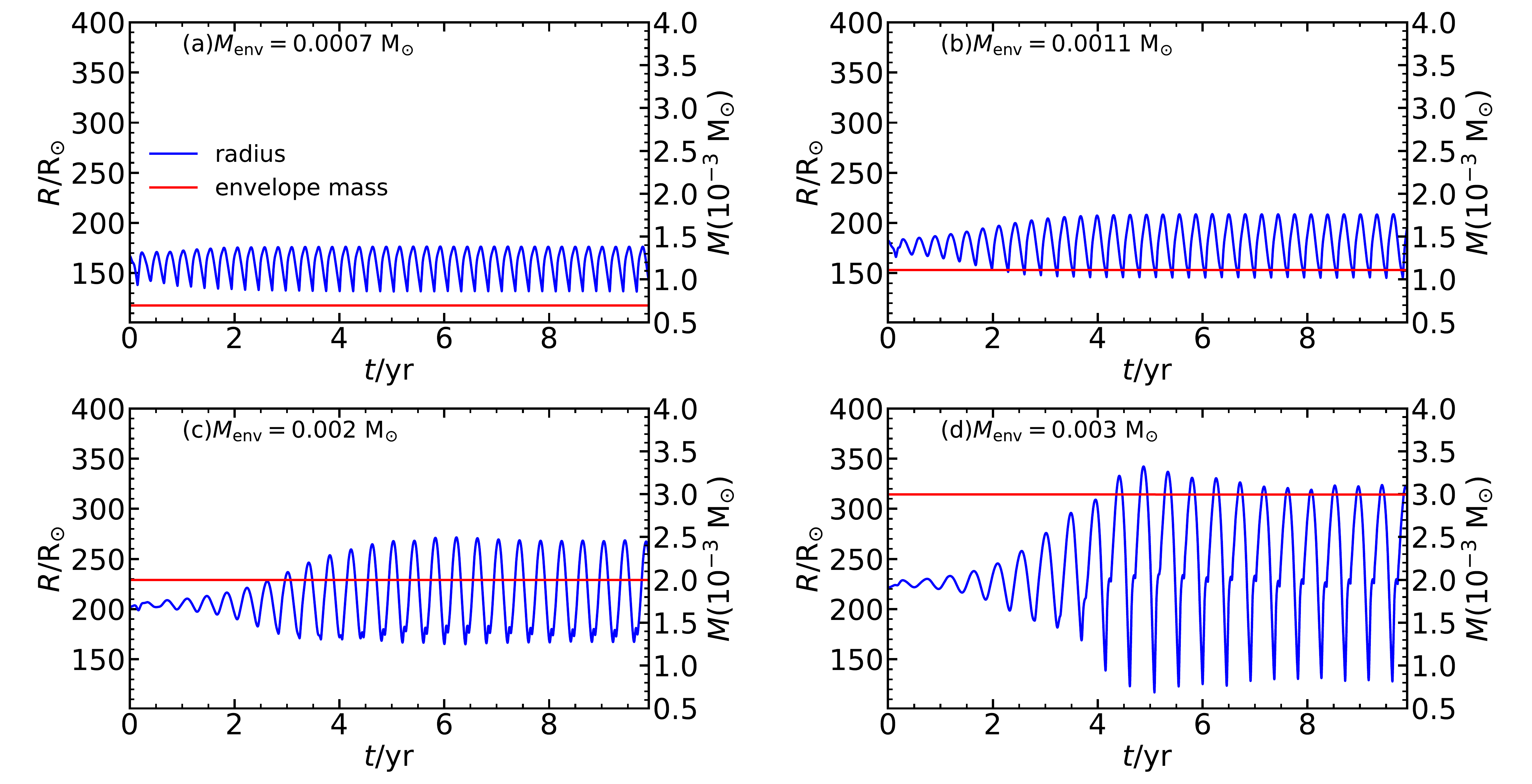}
        \caption{Eevolution of the surface radius (blue) and the envelope mass (red) from a series of models with hydrodynamical effects included. Each panel shows a model for a given initial envelope mass, where the range of envelope mass is  $0.0007\,\mathrm{M}_\sun$\,--\,$0.003\,\mathrm{M}_\sun$. }
        \label{radius}
\end{figure*}

\begin{figure*}
        \centering
        \includegraphics[width=17cm]{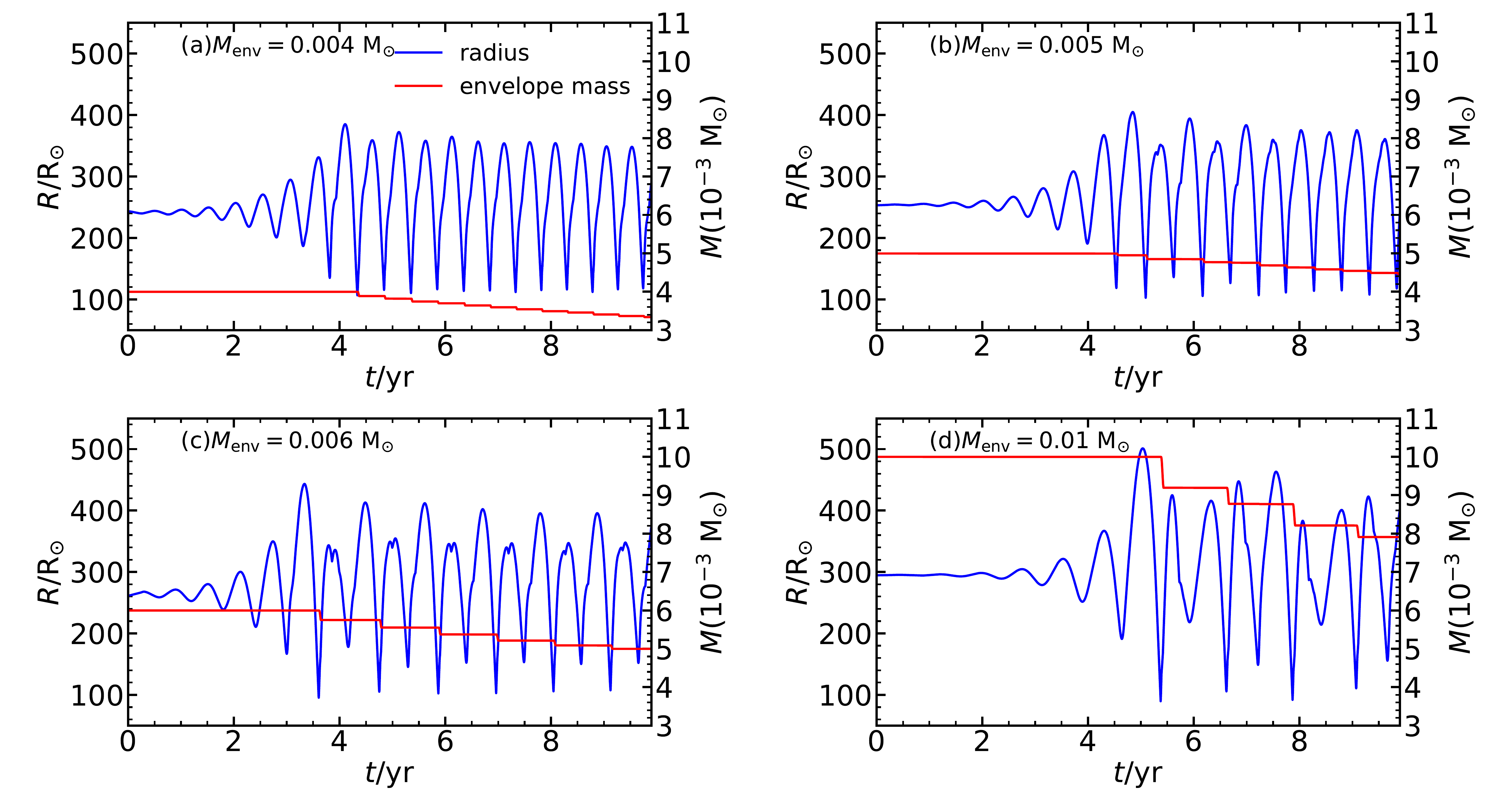}
        \caption{Evolution of the surface radius (blue) and the envelope mass (red) for a series of models with hydrodynamical effects included, similar to Fig.~\ref{radius}, but for larger initial envelope masses. Each panel shows a model with a given initial envelope mass, where the envelope masses range from $0.004\,\mathrm{M}_\sun$ to $0.01\,\mathrm{M}_\sun$. The change of the envelope mass reflects the fact that, once the outer layers exceed the escape velocity of the system, they are removed.}
        \label{radius2}
\end{figure*}

\begin{figure}
        \centering
        \resizebox{\hsize}{!}{\includegraphics{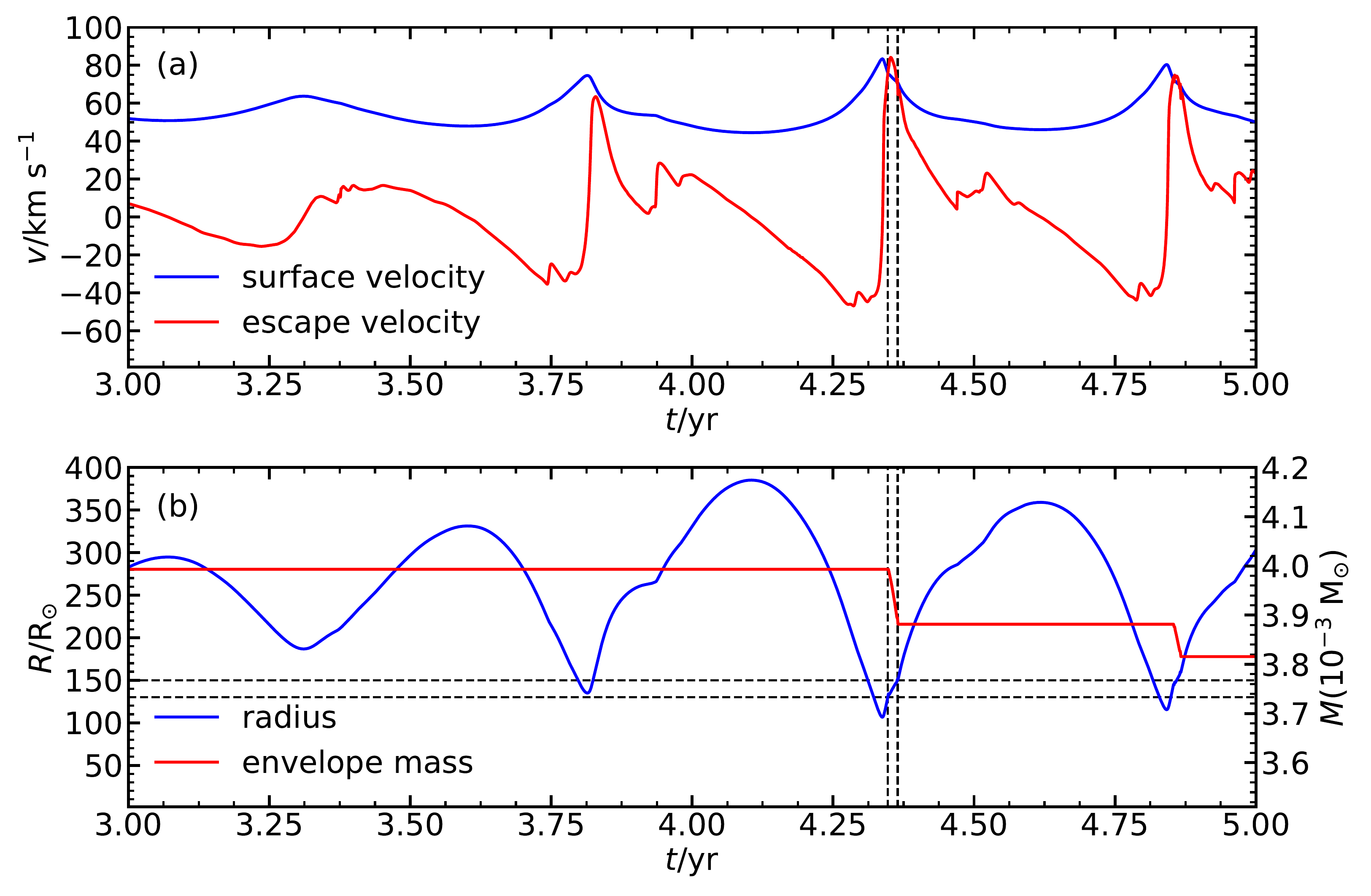}}
        \caption{Example of a hydrodynamical simulation with an envelope of $0.004\,\mathrm{M}_\sun$ from the amplitude increasing phase to the second mass-ejection phase. Panel (a) shows the evolution of the escape velocity at the surface (blue) and the surface velocity of the star (red). Panel (b) shows the evolution of the surface radius (blue) and the envelope mass (red). The dashed lines give the location from where the unbound material is ejected.}
        \label{velocity}
\end{figure}

\subsection{Results for hydrodynamical simulations}
Figure~1(b) shows the evolution of the surface radius and the envelope mass of a system with an envelope of 0.06$ \,\mathrm{M}_\sun  $. The system experiences roughly three major pulsational phases with large associated mass loss: (i) an amplitude-increasing phase, in which,  for a small perturbation, the system starts to pulsate and the amplitude of the pulsation increases gradually until the surface velocity exceeds the local escape velocity, at which point mass ejection occurs; (ii) a mass-loss phase, in which, at the early mass-ejection stage, the mass-loss rate is very high, and one pulse may eject as much as $0.03\,\mathrm{M}_\sun$. As the envelope mass decreases, the pulsation amplitude and the mass-loss rate also gradually decrease; and (iii) a no mass-loss phase, where when the envelope mass decreases below about 0.003$\,\mathrm{M}_\sun $, dynamical mass ejection stops.

As the behavior of the pulsations changes with decreasing envelope mass, we now consider different initial envelope masses. Figures~2 and 3 show the evolution of the surface radius and the envelope mass of the models with an initial envelope of 0.0007$\,\mathrm{M}_\sun $ to 0.01$\,\mathrm{M}_\sun $. In all cases, the hydrostatic-equilibrium solutions are unstable due to small perturbations, and pulsations occur with growing amplitudes. The peak amplitude of the pulsations depends on the envelope mass (i.e., the more massive the envelope, the larger the amplitude). After the amplitude reaches its maximum value for a given system, the subsequent evolution of the system becomes quite different, depending on the initial envelope mass. The models may be divided into two classes, those with and those without mass ejection. Mass ejection always occurs for systems with an envelope more massive than 0.004$ \,\mathrm{M}_\sun $ (see Fig.~3), while it does not occur for systems with an envelope less massive than 0.003$ \,\mathrm{M}_\sun $ (see Fig.~2). We find that for all cases considered here, the pulsations reach their peak amplitudes within 5 yr after the start of the dynamical evolution. Here we only exhibit the evolution within 10 yr. The subsequent long-term evolution is similar to that shown in Fig.~\ref{different methods}(b), that is, if the initial envelope mass is larger than 0.003$ \,\mathrm{M}_\sun $, mass ejection events will reduce it to about 0.003$ \,\mathrm{M}_\sun $; if the initial mass is lower than 0.003$ \,\mathrm{M}_\sun $, the properties of the pulsations will not change significantly after the pulsation amplitude reaches its peak.

Now we want to check whether the occurrence of mass ejection events during the pulsations depends on the amplitude of the previous pulsation. Figure~\ref{velocity}(a) shows the evolution of the surface velocity and the escape velocity for the case with an envelope of $ 0.004\,\mathrm{M}_\sun $, starting with the amplitude-increasing phase to the second mass ejection. Figure~\ref{velocity}(b) presents the corresponding evolution of the radius and the envelope mass. The figures show that after the envelope reaches its minimum radius, it experiences a strong rebound and the surface velocity is accelerated to its maximum value almost instantaneously (as can be seen at about 3.85, 4.35, and 4.85 yr). There is a tendency that, the larger the amplitude of the previous pulsation, the larger the maximum velocity during the following rebound. The reason is that a larger amplitude implies that there is a larger amount of oscillation energy in the envelope, and therefore the envelope will gain more kinetic energy during the next expansion phase. If the amplitude is large enough, the following rebound will be sufficiently strong to accelerate the outer layer to a velocity exceeding the surface escape velocity (i.e., it will lead to mass ejection).

\subsection{Mass loss}
\begin{figure} 
        \resizebox{\hsize}{!}{\includegraphics{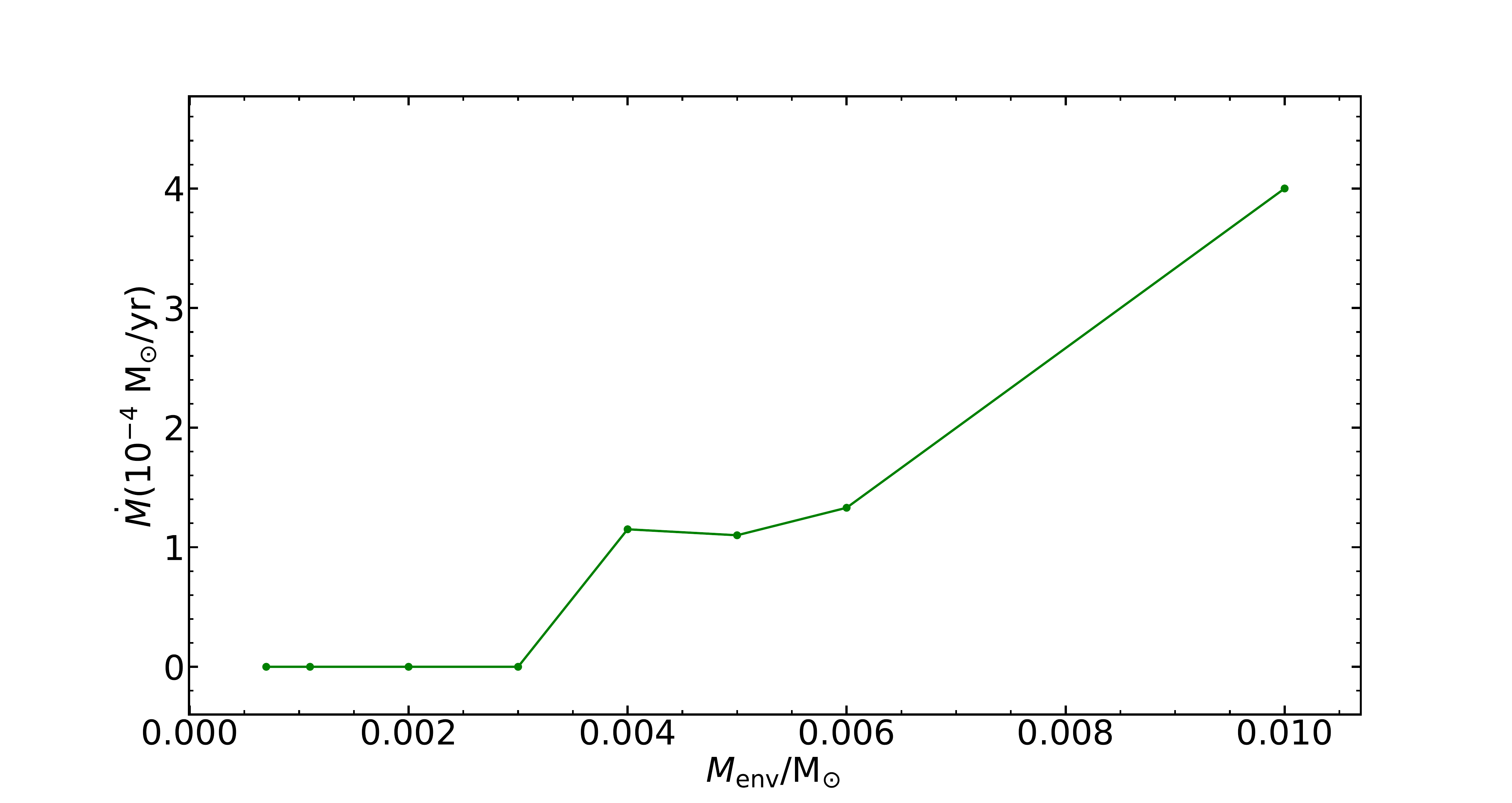}}
        \caption{Relationship between the average mass-loss rate and the envelope mass, where the average mass-loss rate is estimated from Figs.~\ref{radius} and \ref{radius2}. }
        \label{<wind>}
\end{figure}

In this section, we discuss how mass loss affects the CE mass in the CEW model. According to \cite{Meng2017MNRAS}, the rate of change of the envelope mass can be expressed as 
\begin{equation}
\dot{M}_{\mathrm{CE}} = \left| \dot{M}_2\right| - \dot{M}_{\rm WD} -\dot{M}_{\rm wind},
\label{massloss}
\end{equation}
where $ \left| \dot{M}_2\right| $ is the mass-transfer rate and is typically $\sim 10^{-6}\,\mathrm{M}_\sun/\rm yr $ in most cases. The parameter $ \dot{M}_{\rm WD} $ is the mass-growth rate of the WD and is equal to the critical accretion rate of the WD \citep{Hachisu1999ApJ...522..487H}, which is therefore smaller than $ \left| \dot{M}_2\right| $ during the CE phase. The parameter $ \dot{M}_{\rm wind} $ is the mass-loss rate from the CE surface and can be estimated from the results of our simulations. 

Figure~\ref{<wind>} shows the effective time-averaged mass-loss rate, $ \dot{M}_{\rm wind} $, as a function of initial envelope mass. Here, the mass-loss rate is estimated by $M_\mathrm{ej}/\Delta t$, where $M_\mathrm{ej}$ is the amount of ejected material during the first eight mass ejection phases and $\Delta t$ is the total duration of these phases (see also Figs~\ref{radius} and \ref{radius2}).  At first glance, $ \dot{M}_{\rm wind} $ depends on the initial envelope mass (i.e., the higher the envelope mass, the higher the mass-loss rate). However, there is also a small peak around 0.004 $ \mathrm{M}_\sun $, which arises from the different ejection pattern for the different envelope masses. If the envelope is less massive than 0.0047 $ \mathrm{M}_\sun $, mass ejection occurs once during every pulse. Otherwise, it occurs once every two pulses, as shown in Fig.~\ref{radius2}(b).

However, even for the case with 0.004 $ \mathrm{M}_\sun $, $ \dot{M}_{\rm wind} $ is as large as about $ 1.1 \times 10^{-4}\,\mathrm{M}_\sun/\rm yr $, which is much larger than $ \left| \dot{M}_2\right| $, indicating that the CE can grow to above 0.003$ \,\mathrm{M}_\sun $ by mass transfer from the companion in the CEW model, but cannot exceed 0.004 $ \mathrm{M}_\sun $ because of the strong mass loss. Therefore, when the envelope reaches a value of about 0.003 $ \mathrm{M}_\sun $, the envelope mass will maintain an almost constant value; then the mass-loss rate from the surface of the CE in the CEW model can be expressed as $ \dot{M}_{\mathrm{wind}}=\left| \dot{M}_2 \right| -\dot{M}_{\mathrm{WD}} $, until the envelope is less massive than 0.003 $ \mathrm{M}_\sun $, when mass ejection stops.

\begin{figure} 
        \resizebox{\hsize}{!}{\includegraphics{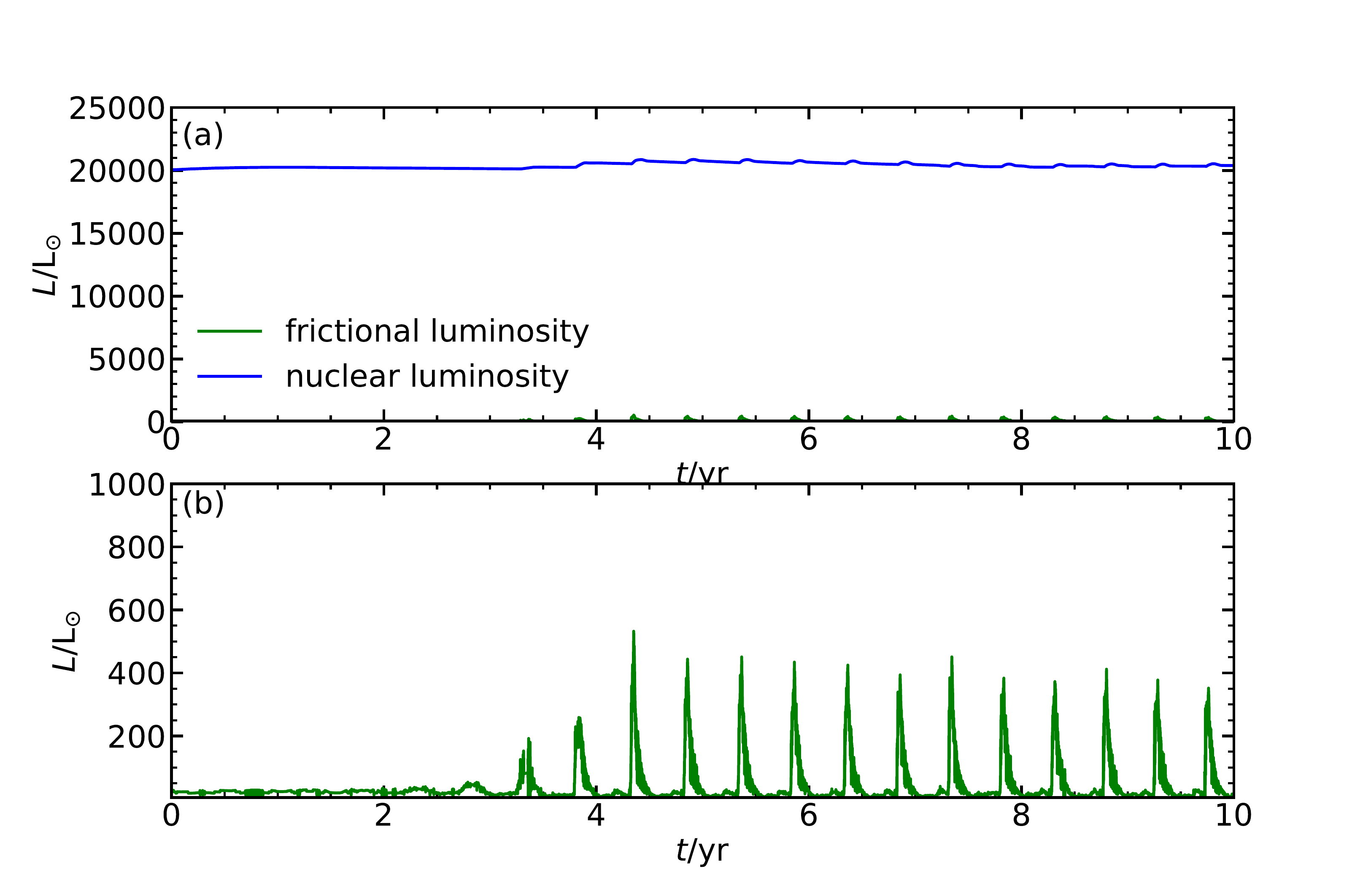}}
        \caption{Evolution of the nuclear luminosity caused by H-shell burning (blue) and the frictional luminosity due to the friction between the binary and the CE (green, from the model with an envelope of 0.004 $ \mathrm{M}_\sun$). Panel (b) shows an enlarged view of the evolution of the frictional luminosity.}
        \label{<friction>}
\end{figure}
\subsection{Frictional luminosity}
\begin{figure} 
        \resizebox{\hsize}{!}{\includegraphics{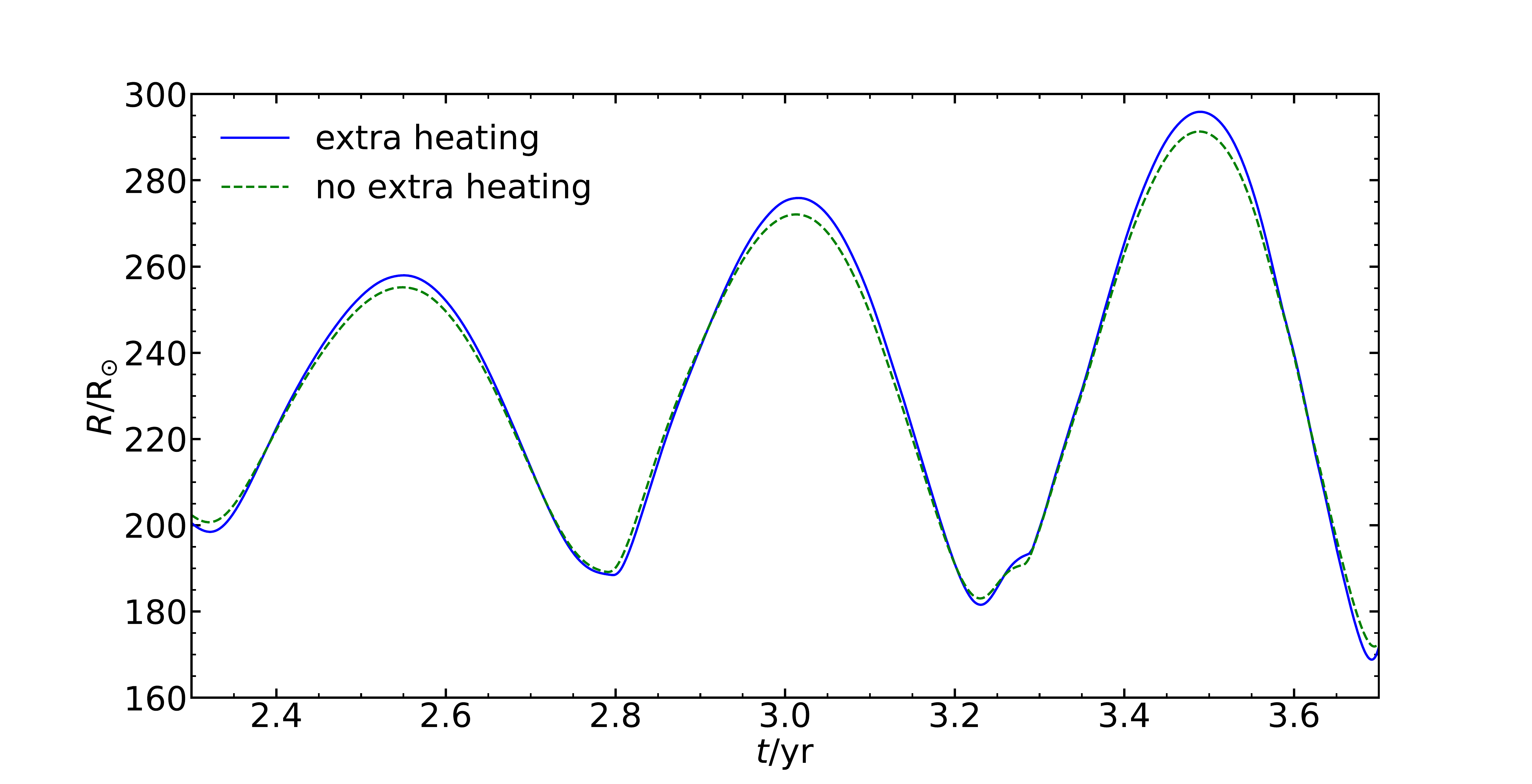}}
        \caption{Evolution of the radius of a model with extra heating (blue line) or without extra heating (dotted green line) sources, where the initial envelope is 0.004 $ \mathrm{M}_\sun$.  }
        \label{<heateffect>}
\end{figure}

During the CE phase, friction between the binary and the CE can affect the system in two ways. One is that the energy generated by the friction causes an expansion of the CE and may lead to its ejection. The other is that the friction extracts orbital energy and thereby reduces the binary separation. In this section, we examine these two effects on the evolution of the CE in the CEW model. 

The extraction rate of the orbital energy by friction can be expressed by a frictional luminosity, which here is calculated as in \cite{Meyer1979A&A}:
\begin{equation}
L_{\rm f} = \alpha \eta G(M_1 + M_2), 
\label{eq frictional}
\end{equation}
where $ M_1 $ and $ M_2 $ are the masses of the WD and the companion, respectively, $ \alpha $ is a coefficient of order one and is set to be\footnote{$\alpha$ is taken as $6\pi$ in \cite{Meyer1979A&A}. However, as we discuss in Sect.~4.1, such a value is not suitable for the low-mass envelopes found in the CEW model.} $\alpha = 1$, and  $ \eta $ is the effective turbulent viscosity,
\begin{equation}
\eta = \rho v_{\rm c} l/2,
\label{viscosity}
\end{equation}
where $ \rho $ and $ v_{\rm c} $ are the local density and the convective velocity near the companion, and $ l $ is the size of the region near to the corotating region where most of the frictional energy is released, which is set to be 10\% of the binary separation following \cite{Meng2017MNRAS}. 

Figure~\ref{<friction>}(a) shows the evolution of the hydrogen shell-burning luminosity and the frictional luminosity for the case with an envelope of $0.004\,\mathrm{M}_\sun$, with an enlarged view of the frictional luminosity in Fig.~\ref{<friction>}(b). The average frictional luminosity is at least two orders of magnitude smaller than the nuclear burning luminosity; this implies that the structure and the evolution of the envelope in the CEW model would not be significantly affected by the frictional heating. To verify this hypothesis, we carried out two simulations,  one with and one without frictional heating sources, where the initial envelope was $0.004\,\mathrm{M}_\sun$. Figure~\ref{<heateffect>} shows the comparison of the evolution of the radius in these two cases: indeed, as expected, the difference is very small. This is also one of the main differences between the CEW model and more typical CE situations where the frictional heating usually dominates the evolution of the CE \citep{Ivanova2013A&ARv..21...59I}.

\begin{figure}
        \resizebox{\hsize}{!}{\includegraphics{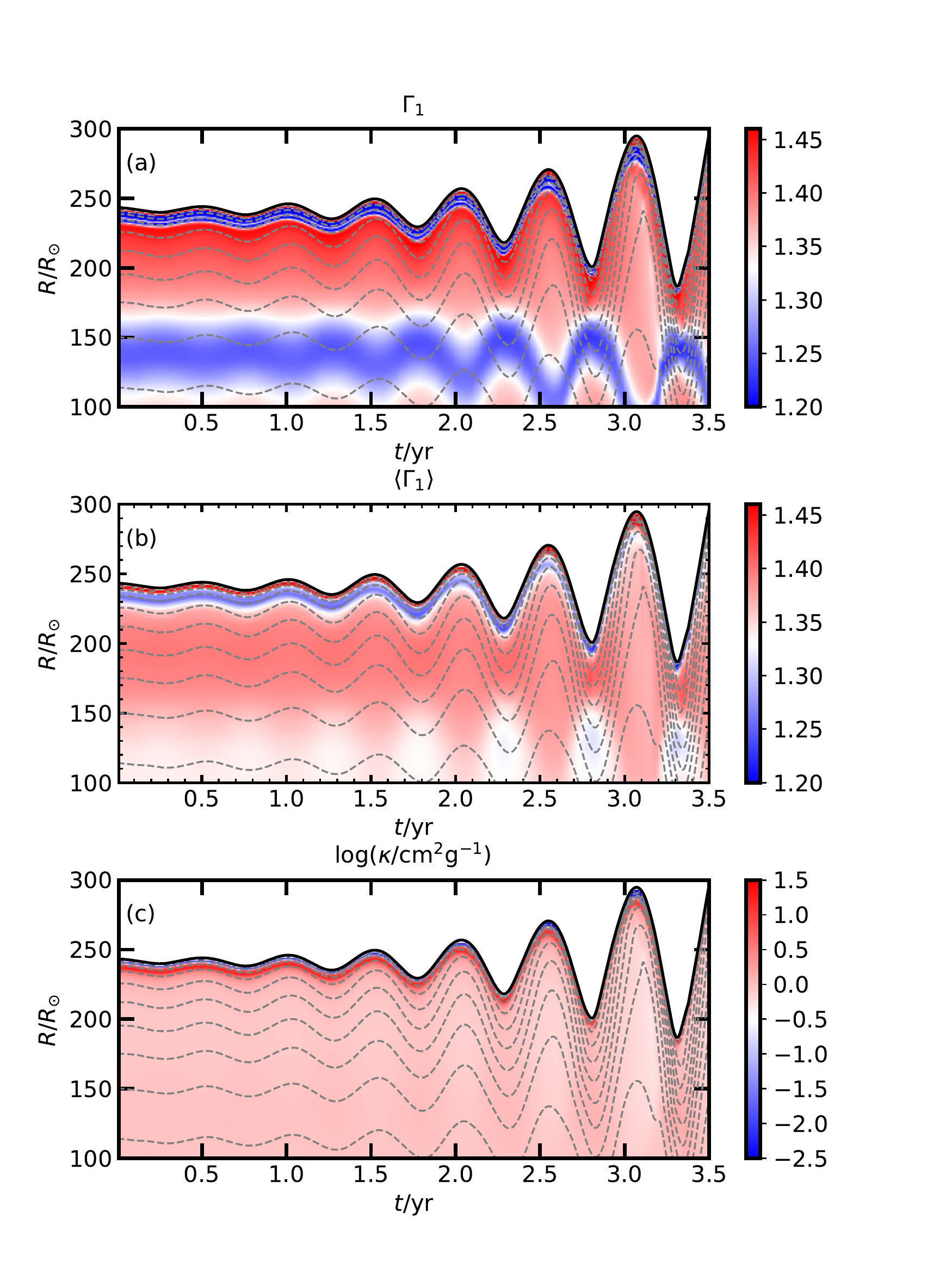}}
        \caption{Amplitude-increasing phase in the model with an envelope of $0.004\,\mathrm{M}_\sun$, showing the first adiabatic exponent, the pressure-weighted volume-averaged value of the first adiabatic exponent, and the opacity. The white color corresponds to the critical value of 4/3 in panels (a) and (b). The dashed lines give the contours containing 90, 80, 70, 60, 50, 40, 30, 20, and 10 percent of the envelope mass.}
        \label{kappa_gamma}
\end{figure}

\begin{figure} 
        \resizebox{\hsize}{!}{\includegraphics{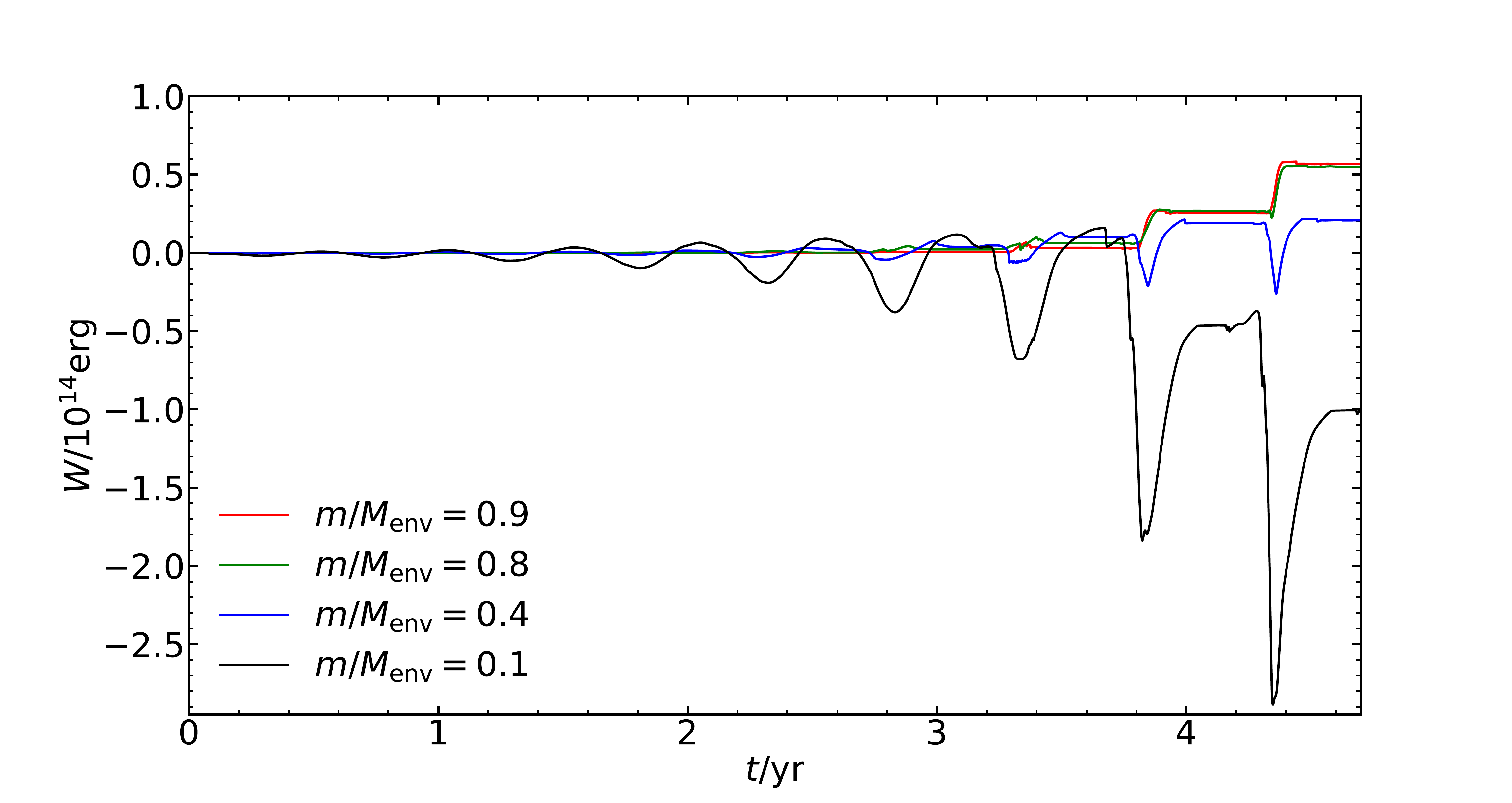}}
        \caption{Cumulative work done by each gram of material in the traced mass shell on its surrounding. The total envelope mass in this case is $0.004\,\mathrm{M}_\sun$, and the four cases shown are for mass shells containing 90, 80, 40, and 10 percent of the envelope mass. The location of these mass shells in the envelope can be seen in Fig.~\ref{kappa_gamma} as dashed lines.  }
        \label{work}
\end{figure}

\subsection{Mechanism for pulsational instability}
A system may oscillate when its equilibrium is perturbed. If the oscillation amplitude grows in time, the system is pulsationally unstable. As shown in Figs.~\ref{radius} and \ref{radius2}, the systems in our simulations are all pulsationally unstable. Generally, there are two types of pulsations: forced pulsations and spontaneous pulsations \citep{Li1992A&A...257..145L}. Forced pulsations are driven by some extra energy source in the envelope. For example, in the CEW model, the frictional heating could be such an energy source. However, as we show in Fig.~\ref{<heateffect>}, even though the heating source is not included in the envelope, the system still begins to pulsate and the pulsation amplitude gradually increases, which indicates that the pulsation in the CEW model is not driven by the frictional heating. The spontaneous pulsation is triggered by some kind of self-exciting mechanism, in which heat energy is transformed into oscillation energy. 

The growth of the oscillation amplitude indicates that some layers in the envelope are dynamically unstable. Dynamical stability is characterized by the first adiabatic exponent $ \Gamma_1 $,
\begin{equation}
\Gamma_1 = \left( \dfrac{\partial \ln P}{\partial \ln \rho}\right)_\mathrm {ad},
\label{adiabitic}
\end{equation}
where $ P $ is pressure and $ \rho $ is density. A value of $ \Gamma_1<4/3 $ in a layer means that a particular layer is dynamically unstable. On the other hand, the pressure-weighted volume-averaged value of $ \Gamma_1 $ reflects the dynamical stability of the whole envelope \citep{Ledoux1945ApJ...102..143L,Stother1999MNRAS.305..365S}, where
\begin{equation}
\left\langle \Gamma_1(m) \right\rangle = \dfrac{\int_{m}^{M_*} \Gamma_1 P\mathrm{d}V}{\int_{m}^{M_*}P\mathrm{d}V},
\label{averaged adiabitic}
\end{equation}
where $ m $ is the mass coordinate and $ M_* $ is the mass of the model in the simulation. $ \left\langle \Gamma_1(m) \right\rangle < 4/3 $ means that material outside the coordinate $ m $ becomes unstable and may be dynamically ejected. Figure~\ref{kappa_gamma} shows how the opacity, $ \Gamma_1 $, and $ \left\langle \Gamma_1(m) \right\rangle $ change with time in the envelope. In Fig.~\ref{kappa_gamma}(a) we can identify  two unstable regions where $ \Gamma_1<4/3 $, one  near the surface, and the other relatively deep inside the envelope. In the unstable surface region, the dynamical instability is driven by the partial ionization of hydrogen and the first partial ionization of helium. The second instability region in Fig.~\ref{kappa_gamma}(a) is driven by the second partial ionization of helium, but this region seems not to be effective in driving the instability of the envelope because it is located too deep inside the envelope (as can be seen in Fig.~\ref{kappa_gamma}(b)).

However, these criteria only allow us to identify regions of dynamical instability in the envelope;  whether they can drive the pulsational instability is still somewhat controversial \citep{Stother1999MNRAS.305..365S}. To further identify the excitation mechanism for the pulsation, we considered the fluid elements in the envelope as cyclic heat engines that do $ P{\rm d}V $ work during the evolution. We used a cumulative work integral $W$ to represent the total net work done by a fluid element on its surroundings.
If a layer does net positive work on average to its surroundings, it tends to drive pulsations. Otherwise, the layer will damp the perturbation. $W$ is defined as
\begin{equation}
W = \int_{0}^{t} P\frac{{\mathrm d} v}{{\mathrm d}\tau}\, {\mathrm d}\tau = \int_{0}^{t} P\frac{1}{\rho}\frac{{\mathrm d}(\ln\rho)}{{\mathrm d}\tau}\, {\mathrm d}\tau,
\label{(total_work)}
\end{equation}
and represents the cumulative work done by the fluid element of unit mass on its surroundings. $ P $ and $v$ are the pressure and the specific volume. We converted the latter into  $ 1/\rho $ for convenience. Figure~\ref{work} shows the evolution of the cumulative work done by the fluid elements of unit mass in four different mass shells, which are defined by their $ m/M_{\mathrm env} $ values, where $m$ gives the envelope mass within a shell and $ M_\mathrm{\mathrm env} $ is the total envelope mass. The locations of these mass shells in the envelope are indicated in Fig.~\ref{kappa_gamma} as dashed lines. As shown in Fig.~\ref{work}, the fluid elements in the layers near the surface do, on average, positive work on their surroundings, while the inner fluid elements do negative work. In the partial ionization zone, the adiabatic compression leads to an opacity increase and vice versa, in which case the work done during a cycle can be positive \citep{Kippenhahn2012sse..book.....K}. The comparison of Fig.~\ref{work} with Fig.~\ref{kappa_gamma}(c) confirms that the pulsation is driven by the layers in the zone of partial ionization of hydrogen and the first partial ionization of helium. Consequently, the pulsation in the CEW model is excited by ionization and recombination, and the excitation mechanism is the so-called $ \kappa $ mechanism.

The $ \kappa $ mechanism can also easily explain why the amplitude of the pulsations increases with increasing initial envelope mass. A system with a more massive initial envelope has a larger radius, and hence a lower effective temperature, in which the characteristic temperature of the ionization zone is found deeper in the envelope where the density is higher. As a result there is more material in the partial ionization zone to drive the pulsation as a heat engine. 

\section{Discussion}
\subsection{Estimate of the spiral-in timescale}
For a WD to explode as a SN Ia, it has to increase its mass to approach the Chandrasekhar mass, which requires a sufficiently long spiral-in timescale of the binary during the CE phase in the CEW model (normally longer than $ 10^5 - 10^6 $ yr). According to \cite{Meng2017MNRAS}, the spiral-in timescale can be estimated by
\begin{equation}
        t_{\rm s} = \dfrac{E_{\rm orbit}}{L_{\rm f}} = \dfrac {GM_{1}M_{2}}{2aL_{\rm f}},
        \label{spiral_in}
\end{equation}
where $E_{\rm orbit}$ is the initial orbital energy, $M_1$ and $M_2$ are the masses of the WD and the companion (here $M_1$ and $M_2$ are 1$\,\mathrm{M}_\sun$), and $a$ is the initial separation of the binary, which is set to be $40\,\mathrm{R}_\sun$ in this paper. From Fig.~\ref{<friction>}(b) in Sect.~3.4, the average value of the frictional luminosity $ L_{\mathrm f} $ is about 100 $ \,\mathrm{L}_\sun $, and therefore the spiral-in timescale is about 4500 yr, which means that the system would merge before the WD could explode as a SN Ia.

The above problem may be caused by the difference between the canonical CE model and the CEW model. Normally, the CE in a CE object forms on the donor's dynamical timescale, which can lead to a very massive CE on a very short timescale. Once the moment of inertia of the CE becomes large enough, corotation with the CE cannot be maintained as a small shrinkage of the orbit and corresponding increase in the orbital angular velocity does not release enough angular momentum from the orbit to accelerate the CE to maintain corotation \citep{Counselman1973ApJ...180..307C,Meyer1979A&A}. After the CE loses corotation with the binary, the strong friction between the binary and the envelope leads to a so-called rapid spiral-in phase, in which the binary orbit shrinks on a dynamical timescale. However, the CE in the CEW model is maintained on the donor's thermal timescale, which is much longer than the dynamical timescale of the accretor. Considering that the mass ejection during the CE phase is so strong, a massive CE cannot be formed.

The above argument can be proved by comparing the moment of inertia of the binary with that of the CE. The moment of inertia of the binary is 
\begin{equation}
I_\mathrm{orbit}=\dfrac {M_{1}M_{2}}{M_{1}+M_{2}} a^2,
\end{equation}
where $M_1$ and $M_2$ are the masses of the WD and the companion, which are both set to  $ 1\,\mathrm{M}_\sun $ in this paper, and $ a=40\,\mathrm{R}_\sun $ is the initial separation of the binary. We then estimate the moment of inertia of an envelope with $ 0.003\,\mathrm{M}_\sun $. In this case, the radius of the CE, $R_\mathrm{CE}$, is set to be $ 250\,\mathrm{R}_\sun $, which is obtained from the results shown in Fig.~\ref{velocity}. The moment of inertia of the envelope can be approximately estimated from\footnote{The coefficient of 2/5 is based on the assumption that the material of the envelope is distributed uniformly in a sphere with radius $R_{\rm CE}$; this is just a conservative upper limit but it is sufficient to prove the argument here. A more realistic estimate (e.g., for red-giant envelopes) may be a factor of two to four smaller.}
\begin{equation}
I_\mathrm{CE} = \dfrac{2}{5}M_\mathrm{CE} R_{\rm CE}^{2},
\label{CEangular}
\end{equation}
where $M_\mathrm{CE}$ is the mass of the envelope and is simply assumed to be equal to $ 0.003\,\mathrm{M}_\sun $. Then, the upper limit of the ratio of the moment of inertia of the envelope to that of the binary is
\begin{equation}
\begin{aligned}
\dfrac{I_\mathrm{CE}}{I_\mathrm{orbit}} \simeq 0.081\times & \dfrac{M_\mathrm{in}}{0.003\,\mathrm{M}_\sun}\dfrac{{1\,\mathrm{M}_\sun\cdot 1\,\mathrm{M}_\sun}}{{M_{1}M_{2}}} \dfrac{{M_{1}+M_{2}}}{{2\,\mathrm{M}_\sun}}\dfrac{\left( 40\,\mathrm{R}_\sun\right) ^2}{a^2}\dfrac{R_{\rm CE}^{2}}{\left( 250\,\mathrm{R}_\sun\right) ^2}.
\end{aligned}
\label{provecorotation}
\end{equation}
This shows that $ I_\mathrm{CE} $ is at least one order of magnitude smaller than $ I_\mathrm{orbit} $, which is mainly due to the low mass of the CE in the CEW model. This result implies that the angular momentum released by a small shrinkage of the binary in the CEW model is much larger than the amount of angular momentum required for the envelope to maintain corotation with the binary. This is completely different from the situation in canonical CE objects with massive envelopes \citep[see the details in Sect.~2 in][]{Meyer1979A&A}. 

For the abovementioned reason, we re-estimated the spiral-in timescale by considering the conservation of angular momentum rather than the actual friction process. According to the simulations in this paper, the CE will maintain a constant mass and a corresponding nearly unchanged angular momentum; in other words, the material entering into the CE will be equal to the material lost from the surface. Thus, we can estimate the angular-momentum loss rate of the orbit using the rate of angular momentum that can be extracted from the orbit by the material transferred to the envelope. Based on \cite{Meyer1979A&A}'s model, we mainly considered the region within a mixing length near the orbit where the material can extract angular momentum from the orbit directly. In order to calculate the amount of extracted angular momentum, we simply assumed that the material near the orbit corotates with the binary.  

It is obvious that the above estimate significantly depends on the actual mass transfer rate in the binary system. This can be determined by detailed binary-evolution calculations. Here, we used parameters from the detailed example given in Fig.~2 in \cite{Meng2017MNRAS} to estimate the spiral-in timescale in the CEW model. We adopted the parameters as the model evolved to $7\times10^5$ yr, with a WD mass and companion mass of $ 1.1\,\mathrm{M}_\sun$ and $1.4\,\mathrm{M}_\sun$ respectively. The typical mass transfer rate was $\sim 1\times10^6\,\mathrm{M}_\sun/\mathrm{yr}$ at this point. Then, the angular momentum taken away by the material transferred into the envelope per unit time could be expressed as
\begin{equation}
\dot{J}=\dot{M}\omega_0 \left( a+{l^{\mathrm{\prime}}}\right)^2,
\end{equation}
where $ \dot{M} $ is the mass transfer rate. $a$ is the binary separation, which is considered to be the boundary of the corotating region \citep{Meyer1979A&A}. $ l^{\mathrm{{\prime}}} $ is the size of the region near to the corotating region where most of the frictional energy is released and was set to $ 0.1\,a\approx 0.61\,\mathrm{R}_\sun $ \citep[for details see][]{Meng2017MNRAS}. Therefore, the spiral-in timescale could be estimated as
\begin{equation}
\begin{aligned}
t_s=\dfrac{J_\mathrm{orbit}}{\dot{J}} \approx 5 \times 10^{5}\ \times &\ \dfrac{{10^{-6}\,\mathrm{M}_\odot \,\mathrm{yr}^{-1}}}{{\dot{M}}}\dfrac{{M_{1}\,M_{2}}}{{1.1\mathrm{M}_\odot\cdot 1.4\mathrm{M}_\odot}}\\&\dfrac{{1.1\,\mathrm{M}_\odot+{1.4\,\mathrm{M}_\odot}}}{{M_{1}+M_{2}}}\dfrac{a^2}{\left( a+{l^{\mathrm{\prime}}}\right)^2}\,\rm yr. 
\end{aligned}
\label{spiral}
\end{equation}

According to the parameters we selected in this article, the typical value of the spiral-in timescale is $5\times10^5$ yr, which is much longer than the 4500 yr estimated previously. This long spiral-in timescale implies that the WDs in the CEW model have, in principle, enough time to increase their masses to $ M_\mathrm{ch} $, and that systems can even survive and recover from a CE phase, as discussed in \cite{Meng2017MNRAS}. In addition, as we discussed in Sect.~3.4, the frictional heating does not affect the evolution of our simulations, and therefore our main conclusions are not sensitive to estimates of the frictional luminosity. We note, however, that the treatment of the spiral-in in a CE system with low-mass envelope remains an important uncertainty in the CEW model, which needs to be addressed in the future. 

\subsection{Shortcoming of the 1D method}
In this work, we use a simplified 1D model that cannot follow the influence of the asymmetrical 3D effects. There are three main 3D effects that can affect the symmetry of the CE: rotation, the companion's gravity, and the frictional heating. Rotation could make the envelope more extended in the equatorial direction, and material in the equatorial region can be more easily ejected than material in the polar direction. This effect might cause a quantitative difference on our results, such as the pulsation properties and the mass-loss rate. However, since the angular velocity is very small near the surface of the envelope, the effect of  rotation on our results is quite insignificant \citep[see Fig.~2 in][]{song2020AA}. The companion's gravity is also unlikely to  significantly affect the symmetry of the envelope because the scale of the envelope is much larger than the separation of the binary in the CEW model. 

Moreover, in our simulations we injected the heating energy uniformly into a spherical shell in the envelope, while in the true 3D model the frictional heating should be more localized. In a canonical CE model, the frictional luminosity is always much larger than the nuclear luminosity inside the system, which can lead to an asymmetry in the envelope. However, the frictional luminosity in the CEW model is too low to affect the envelope structure.

\subsection{Observational signature}
\begin{figure} 
    \resizebox{\hsize}{!}{\includegraphics{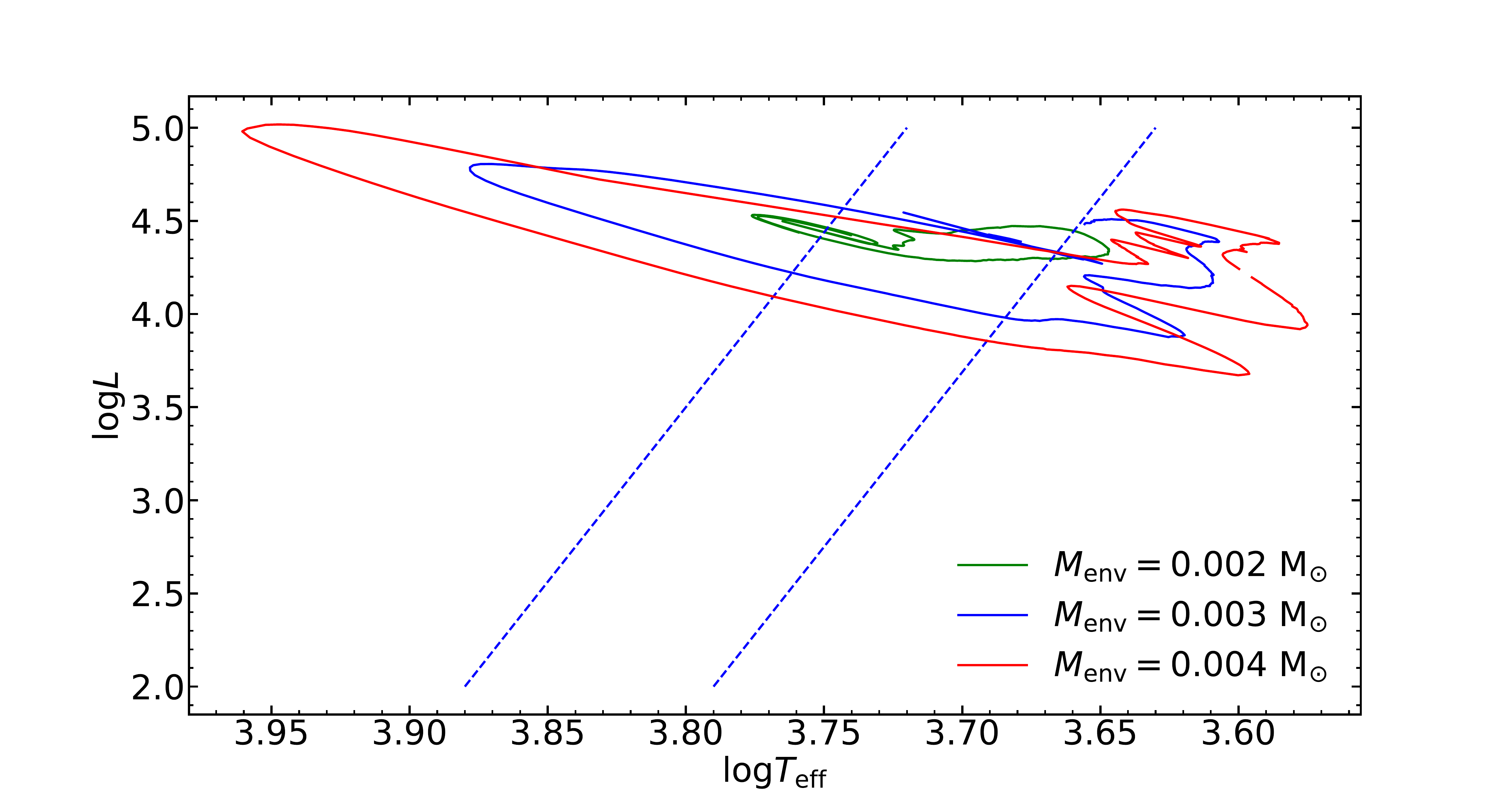}}
        \caption{Hertzsprung-Russell track of one cycle for models with an envelope of $0.002\,\mathrm{M}_\sun$ (green), $0.003\,\mathrm{M}_\sun$ (blue) and $0.004\,\mathrm{M}_\sun$ (red). The dashed blue lines show the boundary of the classical Cepheid instability strip.}
        \label{<HR>}
\end{figure}

According to our simulations, we can make some predictions about the appearance of the systems during the CEW phase. In this paper, we found that the envelope mass cannot exceed 0.004$ \,\mathrm{M}_\sun $ because of the strong mass loss; this provides strong constraints on the observational properties of the systems. 

First, these systems are experiencing periodic pulsations, and thus should appear as variable stars. The period increases with the envelope mass and is in the range of 0.25-0.5 yr, and the amplitude of the brightness variation may be as large as 2.3 magnitude. 

Second, the ejected material can form a CSM around the system. The typical ejecta velocity is about $ 50 - 80\,\mathrm {km/s} $, which is consistent with the constraints from observations for SNe Ia with inferred CSM \citep{Patat2007Sci...317..924P,Dilday2012Sci...337..942D}.

Based on the surface parameters of the systems from our simulations, we can estimate their location in the Hertzsprung-Russell (HR) diagram. The HR tracks of one pulsation from the models with the envelope of 0.002, 0.003, and 0.004$ \,\mathrm{M}_\sun $ are shown in Fig.~\ref{<HR>}. In the HR diagram, the central point of $ \log L $ lies between 4.2 and 4.4 and of $ \log T_{\rm eff} $ between 3.7 and 3.8, meaning they are located in the classical Cepheid instability strip, also shown in the figure, which again confirms our conclusion that the excitation mechanism in our simulations is indeed the $ \kappa $ mechanism, just as in classical Cepheids. 

Because the CE masses of the system are quite low, the H-rich material of the CE may not be directly detected in the spectra of SNe Ia. However, such low-mass H-rich envelopes may be essential for understanding the high-velocity features detected in the spectra of SNe Ia \citep{Mazzali2005MNRAS.357..200M}. \cite{Meng2019ApJ...886...58M} found that the strength of the high-velocity feature is related to the stellar population, which favors the interaction between supernova ejecta and a H-rich CSM as the origin of the high-velocity feature. Mass loss from the low-mass H-rich envelopes in the CEW model may be a natural cause for the origin
of the H-rich CSM.

\subsection{Comparison with other classes of pulsating stars}
The structure of the CEW model should be similar to the envelope of a TPAGB star \citep{Meng2017MNRAS,song2020AA}. TPAGB stars can also undergo strong pulsations and are then known as Mira variables \citep{Aerts2010aste.book.....A}. However, the location of a system for the CEW model in the HR diagram is at a higher effective temperature than in typical asymptotic giant branch (AGB) stars, and closer to the Cepheid instability strip due to the additional gravity of the secondary and the rather low mass of the envelope, whereas Cepheids have a rather different structure \citep{song2020AA}. A better comparison can perhaps be made with post-AGB stars, which also have a very low-mass, low-density envelope. The post-AGB stars located in the instability strip are known as RV Tauri variables. They also undergo dynamical pulsations, but typically have much lower luminosity than what is seen in our CEW models \citep{Fokin1994A&A...292..133F,vanWinckel2003ARA&A..41..391V,Giridhar2020JApA...41...44G}. In addition, one of the most characteristic aspects of our results is that the strong pulsations can trigger dynamical mass ejection events, which are not necessarily seen for these other pulsators.

\subsection{Comparison with the OTW model}
In Sect.~1, we introduced the two main difficulties encountered by the OTW model from observations. However, based on our simulations, these two problems can be naturally resolved in the CEW model. One problem is that the outflow velocity from the progenitor system of SNe Ia should not exceed $ 200\,\mathrm {km/s} $ \citep{Badenes2007ApJ}. In Fig.~\ref{velocity}, the velocity of the ejected material is just slightly larger than the escape velocity, namely about $ 50 - 80\,\mathrm {km/s} $, which is therefore consistent with observations. In addition, the ejecta velocity in this paper is also consistent with the estimates in \cite{Meng2017MNRAS}. 

The other problem is that the OTW model has difficulty in explaining the production of SNe Ia in low-metallicity environments  because the OTW model strongly depends on metallicity. When $Z\leq 0.002$, an OTW cannot occur, and hence no SNe Ia are expected in low-metallicity environments. However, in the CEW model, the mass ejection is mainly driven by the ionization zones of H and He, and hence the CEW model does not depend on metallicity. Thus, the CEW model can naturally explain SNe Ia discovered in low-metallicity environments. 

\section{Conclusions}
In this paper, we carried out a series of 1D hydrodynamical simulations to simulate the properties of systems with low-mass envelopes, as expected in the CEW model. The main conclusions are the following:

\begin{enumerate}[1)]
\item In our simulations, all systems are pulsationally unstable, independent of the initial CE mass. As the pulsational amplitude increases, the outer layers of the CE in some systems exceed their surface escape velocity, leading to mass ejection. Whether or not the ejection occurs is heavily dependent on the initial CE mass; in other words, mass ejection only occurs when the initial CE is more massive than 0.003$ \,\mathrm{M}_\sun $. However, the mass-loss rate from systems where the initial CE mass is larger than 0.004$ \,\mathrm{M}_\sun $ is much larger than the mass-transfer rate between the binary components.
This implies that the CE mass in the CEW model cannot significantly exceed 0.003$ \,\mathrm{M}_\sun $. These results do not depend on whether the frictional luminosity between the binary and the CE is included or not. 
\item The $\kappa$ mechanism caused by H and He ionization drives the pulsations. The central points of the evolutionary tracks of the systems in the HR diagram are located in the classical Cepheid instability strip, and the systems appear as periodic variable stars.
\item Using arguments based on the conservation of angular momentum, but which ignore the possible role of friction in a differentially rotating envelope, we roughly estimated that the spiral-in timescale of the binary system in the low-mass CE could be longer than a few $10^5$ yr, which means that the WDs in the CEW model can survive from the CE phase and have enough time to increase their masses to $ M_\mathrm{ch} $ and then explode as SNe Ia. 
        
\end{enumerate}

As a final caveat we mention that, in this paper, our models do not include thermal pulses. Whether the He flash in such low-mass envelopes may destroy the CE or not is currently unclear and will be addressed in a future publication. 

\begin{acknowledgements}
We are grateful to the anonymous referee for his/her constructive comments that helps us to improve the manuscript greatly. This work was supported by the National Key R\&D Program of China with No. 2021YFA1600403, the NSFC (Nos. 11973080 and 11733008). We acknowledge science research grants from the China Manned Space Project, no. CMS- CSST-2021-B07. X.M. acknowledges the support by the Yunnan Ten Thousand Talents Plan Young \& Elite Talents Project, and CAS `Light of West China' Program.
\end{acknowledgements}

\bibliographystyle{aa} % style aa.bst 
\bibliography{ref} % your references Yourfile.bib
\end{document}